\documentclass[12pt]{article}
\usepackage[margin=1in]{geometry}
\usepackage{graphicx}
\usepackage{listings}
\usepackage{amsmath}

\usepackage[
backend=biber,
style=numeric,
sorting=ynt
]{biblatex}

\title{\large {\bf A DATA-DRIVEN APPROACH FOR MODELING THE BEHAVIOR OF STOCK PRICES}}
\author{Khalid Y. Aram \\ \small{\textit{Department of Business Administration, Emporia State University, Emporia, Kansas, USA}}\\ \small{karam@emporia.edu}}
\date{August 2022}

\begin{document}

\maketitle
\section*{ABSTRACT}
In this paper, we describe two approaches to model the behavior of stock prices. The first approach considers the underlying probability distribution of day-to-day price differences. The second approach models the movement of the price as a stochastic birth-death process.  We demonstrated the two approaches using historical opening prices of Apple inc. and compared the simulated prices from the two approaches to the actual ones using information theory metrics.
\section{Introduction}
The term ''behavior of stock prices'' was first introduced in 1965 by Fama \cite{aa}. The definition of this term according to the author involves two aspects: (1) successive price changes according to a stochastic process, (2) and price changes according to some probability distribution. Stock price behavior can be described using mathematical models which can be utilized by decision-makers to inform the decisions of buying or selling stock shares. The problem of mathematically modeling and forecasting stock prices is a tough challenge for finance and financial engineering due to the complex behavior of stock prices. The behavior complexity of stock prices stems from the complexities of other factors that impact the stocks such as supply and demand, market conditions, the media, and others \cite{bb}. The literature contains many approaches to model stock price behavior. One of the popular approaches is to model the movement of stock prices using Markov Chains \cite{cc}. 
The objective of this study is to model the behavior of stock prices using two approaches. The first approach generates prices by sampling from the underlying probability distribution of the one-step price movement. The second is to model the price as a stochastic birth-death process, which is a special case of continuous-time Markov chains. We compared the simulated output of the two models to the actual data using information entropy and mutual information measures. 

\subsection{Birth-Death Processes}
A birth and death process is a special case of continuous-time Markov chains, which have infinite state space. If the system is currently at state \emph{n}, the system has two possibilities: moving forward to state \emph{n}+1, or backward to state \emph{n}-1. Figure 1 is an illustration of a birth-death process. The rate at which the system moves forward is called birth rate $\lambda$, and for moving backward, there is a death rate $\mu$. The time between two state changes is assumed to follow an Exponential distribution with a rate of $\lambda+\mu$. The probability of a birth or a death move can be written as follows \cite{ee}:
\begin{center}
\includegraphics[width=2 in]{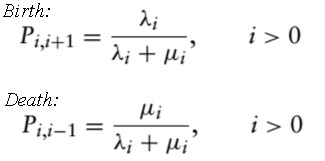}
\end{center}

\begin{figure}
\centering
\includegraphics[width=5 in]{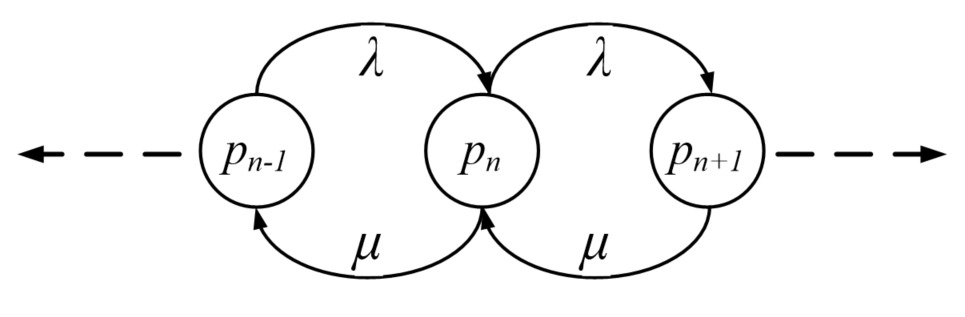}
\caption{Birth-death process. $p_n$ = stock price.}
\end{figure}

\section{Modeling Stock Price Behavior}
This section describes two approaches to modeling the behavior of stock prices. The section also describes the algorithms used to simulate the two models and generate prices. 
\subsection{Stock Price Data}
We used historical stock price data of Apple Inc. from 2016 (251 days) \cite{dd}. We only used the opening prices for modeling and simulation. Figure 2 shows a plot of the historical opening prices. The day-to-day price differences (moves) were calculated. This is simply the difference between prices in two consecutive days: $d_i = p_i - p_{i-1}$. The distribution of day-to-day price differences was used to build the two models in this study.

\begin{figure}
\label{actualprices}
\centering
\includegraphics[width=5 in]{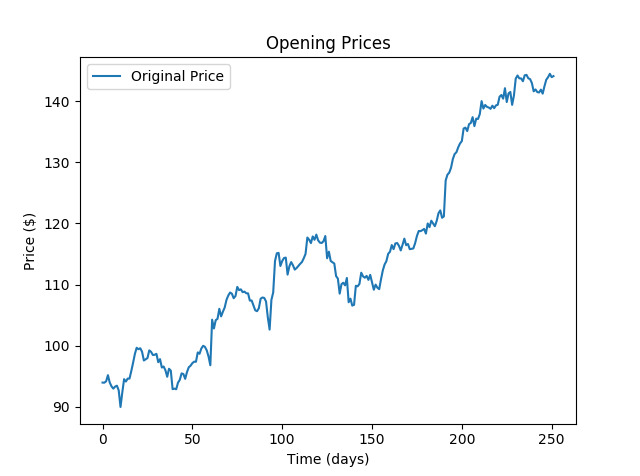}
\caption{One-year historical opening prices for Apple Inc.}
\end{figure}

\subsection{Model 1: Normally Distributed Price Differences}
\begin{figure}[ht!]
\label{actualdiff}
\centering
\includegraphics[width=\columnwidth]{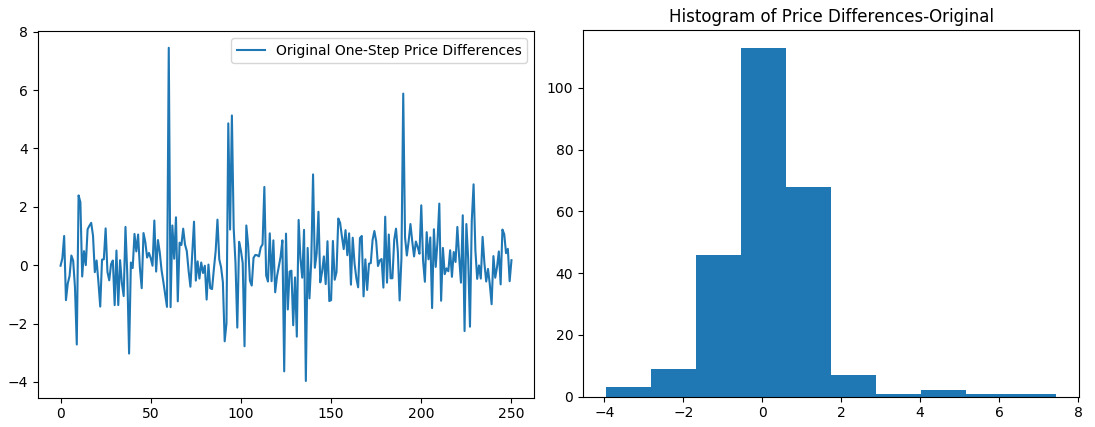}
\caption{A plot and a histogram of price differences for the historical data.}
\end{figure}
Figure 3 shows the day-to-day price differences for the opening prices in 251 days and a histogram of the differences. From the histogram, we can assume that the price difference is normally distributed. The estimated parameters of this distribution are $\mu = \$ 0.1989, \sigma = \$ 1.2782$.

Using the parameters of the assumed distribution, we can randomly generate price moves. If we start from the first opening price in the data (beginning of the year), and stagger up the generated price moves, we can obtain a simulated price trajectory. The following is a pseudo code for the algorithm used to simulate prices. The simulation was implemented in Python 2.7 \cite{ff}.
\begin{lstlisting}
mu = 0.1989
sigma = 1.2782
p = starting price
time = 251  # one year
t = 0
while t < time, repeat:
	generate random number r
	p = p + (mu + sigma * Z(r))  # Z is the standard 
	normal distribution
	observe p
\end{lstlisting}
\begin{figure}[ht]
\label{simnormal}
\centering
\includegraphics[width=\columnwidth]{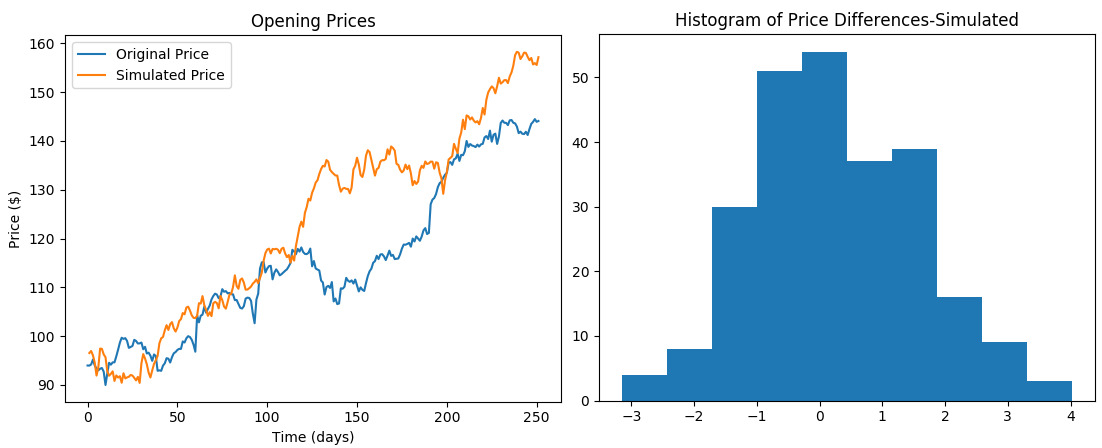}
\caption{A plot of simulated and actual prices and a histogram of model 1 simulated price moves.}
\end{figure}
\begin{figure}[ht!]
\label{100simnormal}
\centering
\includegraphics[width=5 in]{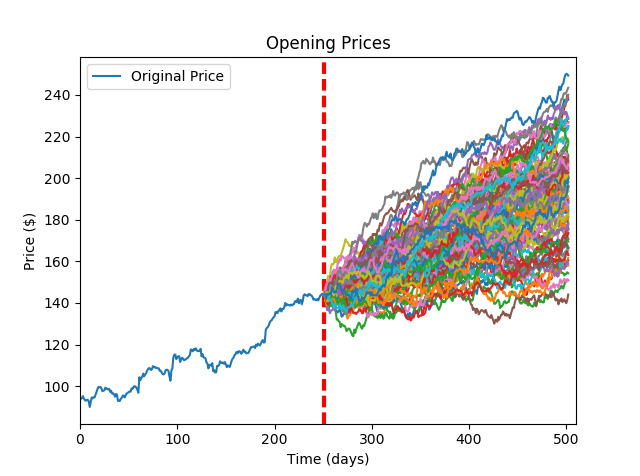}
\caption{100 one-year simulated price trajectories generated by model 1.}
\end{figure}
Figure 4 shows the simulation results. The histogram in the figure is for the generated price moves. In this model, we assume that the distribution of price differences will not change over time, which is not necessarily true. Figure 5 shows 100 simulated trajectories of the opening price for one year ahead. We here assumed that the behavior of the price will be the same for the next year, and this is shown by the increasing trend in the prices, which reflects the same trend in the actual data.

\subsection{Model 2: Continuous-time Birth-Death Process}
In this model, we capture more details of the stock behavior. We model the change in prices as a Markov chain with two states: price increase and price decrease. We also assume that the time between two increases and the time between two decreases are continuous random variables. Thus, this process can be described as a stochastic birth-death process. From the price moves in the actual data, we assume that a positive difference represents an increase (birth), and a negative difference represents a decrease (death). We obtained data for the time between two increases and the time between two decreases and plotted a histogram for each as shown in figure 6 in order to determine the birth rate $\lambda$ and the death rate $\mu$. From the plots, we can see that the two variables are approximately exponentially distributed. We estimated the birth and death rates as $\lambda = 0.5739$ and $\mu = 0.4223$. In other words, the mean time between increases is 1.7 days, and the mean time between increases is 2.3 days.
\begin{figure}[ht!] 
\label{histtime}
\centering
\includegraphics[width=\columnwidth]{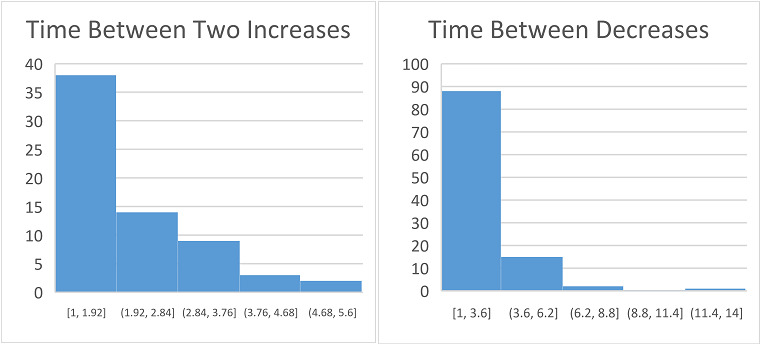}
\caption{Histograms of the time between two price increases and two price decreases.}
\end{figure}

The magnitude of price increment or decrement was also analyzed to obtain probability distributions for each. As shown in figure 7, the price increments and decrements are approximately exponentially distributed with means of \$ 1.0897 and \$ 1.2235, respectively.
The following is a pseudo code for the algorithm used to simulate model 2 and generate price trajectories based on the distributions obtained from the data.
\begin{lstlisting}
Lambda = 0.57
mu = 0.42
p = starting price
time = 251  # one year
t = 0
while t<time, repeat:
	generate random number r
	if r <= lambda/(lambda+mu)
		draw a number i from exponential distribution 
        with mean 1.08
		p = p + i
	else:
		draw a number (d) from exponential distribution 
        with mean 1.22
		p = p - d
	observe p
	t  = t - [log(r)/(lambda + mu)]
\end{lstlisting}
\begin{figure}[ht!]
\centering
\includegraphics[width=\columnwidth]{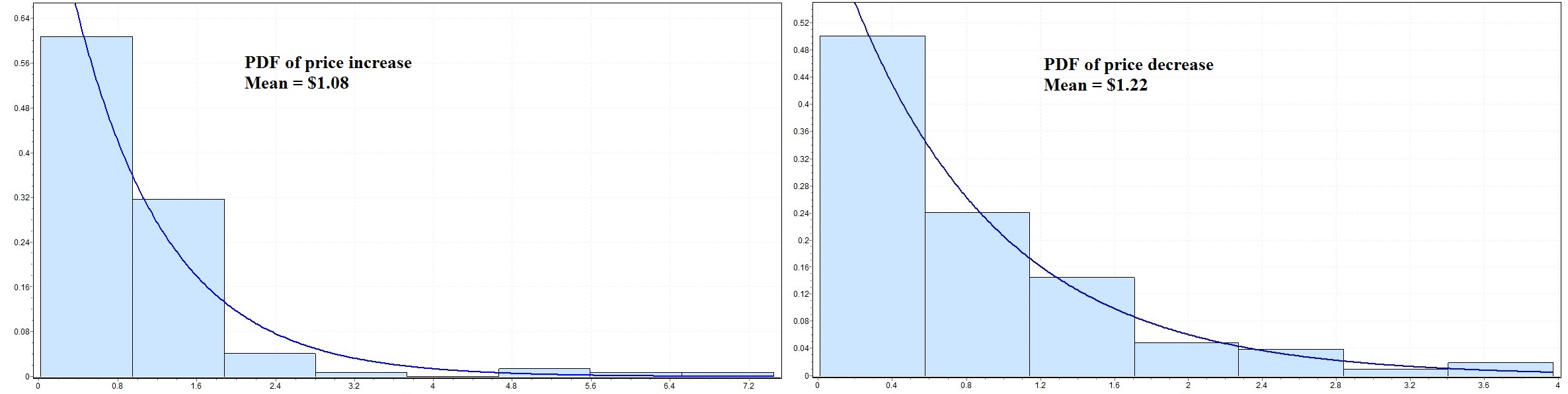}
\caption{Histograms of price increments and decrements.}
\end{figure}
Figure 8 shows simulation results including a histogram of price moves generated by the model. We notice that the histogram is quite similar to that of the actual data in figure 3. Figure 9 shows a plot of 100 predicted future trajectories of the price, which does not have the same trend of the actual data, as that generated by model 1.
\begin{figure} 
\centering
\includegraphics[width=\columnwidth]{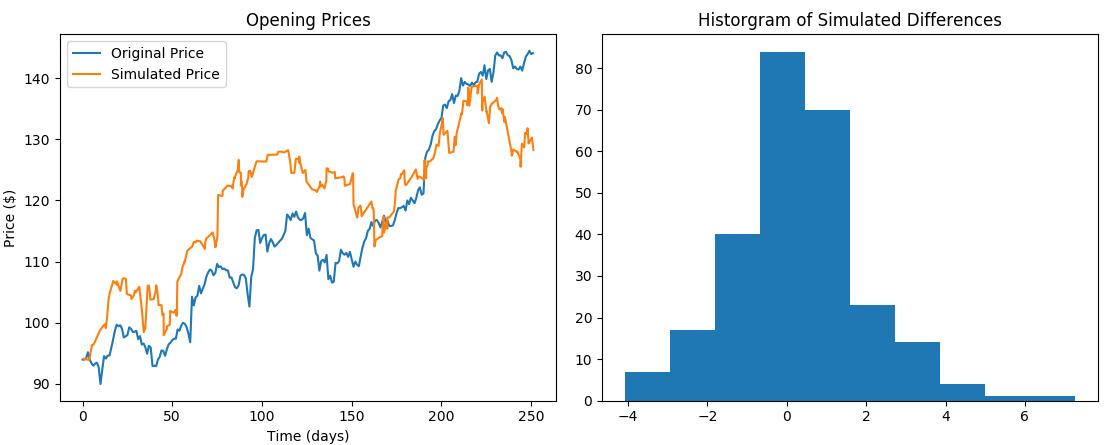}
\caption{A plot of simulated and actual prices and a histogram of model 2 simulated price moves.}
\end{figure}
\begin{figure} 
\centering
\includegraphics[width=5 in]{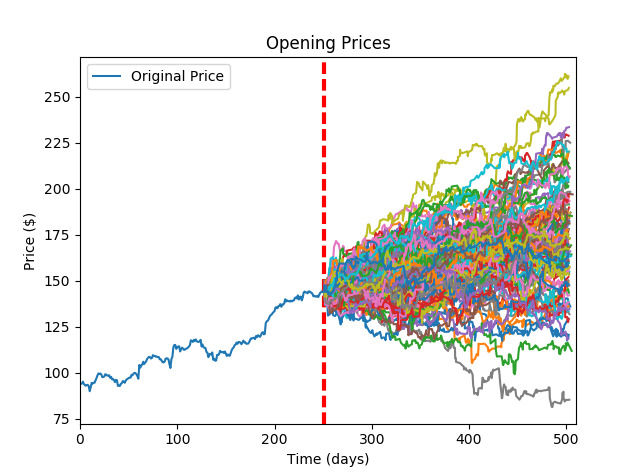}
\caption{100 one-year simulated price trajectories generated by model 2.}
\end{figure}
\section{Evaluating Models Using Information Theory}
We calculate information entropy and mutual information (MI) between prices in two consecutive days for the prices obtained by the two models, and we compare them to those of the historical data. The distribution of price differences is a continuous random variable, so we will use the same histogram bin size for calculating entropy and mutual information for actual and simulated distributions. We set the bin size to 10 and used \emph{base} = 2 to measure entropy and MI in bits. For each measure, an average of 100 replicates was taken. Table 1 shows the evaluation results.

\begin{table}[h]
\caption{Information entropy and mutual information for the two models.}
\centering
\begin{tabular}{ |p{2 in}|p{1.5 in}|p{2.5 in}| } 
 \hline
 {\bf Data} & {\bf Information Entropy} & {\bf Mutual Information for the Price in Two Days} \\ 
 \hline
 Actual Historical Prices & 3.0241 & 1.6213\\ 
 Model 1 Output	& 3.0438 &	1.5758 \\ 
 Model 2 Output &	3.0551 & 1.4066\\
 \hline
\end{tabular}
\end{table}

By comparing the results in table 1, we notice that the two models have information entropy and MI close to those of the actual data. The entropy values of the models are slightly higher, and it is higher for model 2 than for model 1. This shows that the amount of randomness in the models is higher. The MI for the two models is slightly less than the actual one, and it is lower for model 2 than for model 1. Model 1 is expected to have the same measure value as the actual data, as it generates prices using the underlying distribution of the data. In other words, model 1 attempts to mimic the historical trend. However, model 2 has a different mechanism to mimic price moves. Overall, model 2 shows good performance in providing information about the actual data.

MI can be used to measure the similarity between two probability distributions. We calculated MI between the actual price moves distribution and the ones generated by the two models. MI between the actual moves and model 1 differences is 0.1242, and for model 2 moves is 0.1334. Hence, we can conclude that model 2 produces price moves that are closely similar to those of the actual data.

\section{Conclusion}
In this study, we attempted to model the behavior of stock prices. We used two approaches for modeling. The first was to generate prices using the underlying probability distribution of day-to-day price moves. The second model captures more details of the process of price movement using the concept of birth-death stochastic processes. We compared the output of the two models to the actual historical data, and we concluded that the second model produces more similar prices to the actual prices for the selected stock data.

\printbibliography

\end{document}